\documentclass[aps,prd,twocolumn,groupedaddress,nofootinbib,longbibliography]{revtex4-1}

\usepackage{amsfonts,amsmath,amssymb}
\usepackage{nicefrac}
\usepackage{graphicx}
\usepackage{color}
\usepackage{soul} 
\usepackage{hyperref}
\usepackage{accents}
\usepackage[normalem]{ulem}
\usepackage{cleveref}

\def\exd{\mathrm{d}}

\def\d{\delta}
\def\e{\epsilon}
\def\o{\omega}

\def\G{\Gamma}

\def\cH{\cal H}

\def\Lambda{\lambda}

\def\bL{\bold{L}}
\def\bsmu{\boldsymbol{\mu}}
\def\bQ{\bold{Q}}
\def\p{\partial}
\def\n{\nabla}

\def\beq{\begin{equation}}
\def\eeq{\end{equation}}
\def\bea{\begin{eqnarray}}
\def\eea{\end{eqnarray}}

\def\nn{\nonumber}

\begin{document}


\title{Higher derivative gravity: field equation as the equation of state}


\author{Ramit Dey}
\email[]{rdey@sissa.it}

\author{Stefano Liberati}
\email[]{liberati@sissa.it}
\affiliation{SISSA, 
Via Bonomea 265, 34136 Trieste, Italy \\
                INFN, Sezione di Trieste, Trieste, Italy}
\author{Arif Mohd}
\email[]{amohd@umd.edu}
\affiliation{Center for Fundamental Physics,\\ University of Maryland, 
                    College Park, Maryland 20742, USA}



\begin{abstract}
  One of the striking features of general relativity is that the Einstein equation is implied by the Clausius relation imposed on a small patch of locally constructed causal horizon. Extension of this thermodynamic derivation of the field equation to more general theories of gravity has been attempted many times in the last two decades.  In particular, equations of motion for minimally coupled higher curvature theories of gravity, but without the derivatives of curvature, have previously been derived using a thermodynamic reasoning.  In that derivation the horizon slices were endowed with an entropy density whose form resembles that of the Noether charge for diffeomorphisms, and was dubbed the Noetheresque entropy. In this paper, we propose a new entropy
 density, closely related to the Noetheresque form, such that the field equation of any  diffeomorphism invariant metric theory of gravity can be derived by imposing the Clausius relation on a small patch of local causal horizon. 
\end{abstract}

\maketitle

\section{Introduction}
\label{sec:intro}

General relativity, and other diffeomorphism invariant theories of gravity, admit special states called black holes whose mechanics is governed by the laws { that are in exact correspondence} to the laws of thermodynamics \cite{Bardeen:1973gs,Wald:1993nt,Iyer:1994ys}. { The expression for the energy and entropy of these states depend upon the theory under consideration and their temperature is given by a  geometric quantity, namely the surface gravity associated to the black hole horizon.} The latter is identified as the temperature by studying quantum field theory on the gravitational background of the black hole \cite{Hawking:1974sw}. 
Classical and quantum dynamics of black holes is widely believed to provide important lessons for understanding the underlying quantum theory of gravity. However, the underlying quantum theory should describe all gravitational macrostates and not merely the black holes. Thus it seems plausible that if we restrict our attention to a region of spacetime small enough (with respect to the curvature scale) such that  the spacetime is ``close to'' Minkowski, and we assume the validity of the Einstein Equivalence Principle~\cite{DiCasola:2013iia}, then locally the state should look like an equilibrium one and a coarse-grained/thermodynamic description of the degrees of freedom contained in that region of spacetime should be possible.   

 About twenty years ago, this chain of reasoning led Jacobson to derive the Einstein equation as the equation of state of these underlying degrees of freedom~\cite{Jacobson:1995ab}. Assuming that the heat flow corresponds to the energy-momentum flux of matter across the Rindler horizon of a local observer, the entropy corresponds to the area of the horizon, and the temperature has the  Unruh value $(=\hbar/2\pi)$, Jacobson showed that the horizon must be dynamical in order for the  Clausius relation $\exd S=\d Q/T$ to hold true, and that its evolution is governed by the Einstein equation.    
 
Jacobson's approach of deriving the gravitational field equation from the Clausius relation has been applied to other theories of gravity \cite{Elizalde:2008pv,Brustein:2009hy,Parikh:2009qs,Padmanabhan:2009ry,Padmanabhan:2011ex,Padmanabhan:2009vy}. In particular, it was applied to $f(R)$ theory \cite{Eling:2006aw,Chirco:2010sw} after deforming the Clausius relation to account for the internal entropy production terms, $\exd S = \exd_i S + \d Q/T$. All these approaches have been critically reviewed by Guedens, Jacobson and Sarkar~\cite{Guedens:2011dy} whose work has inspired our study. 
 The authors of ref.~\cite{Guedens:2011dy} used a careful construction of the geometry of local causal horizon (LCH)  and the approximate Killing vector field constructed in ref.~\cite{Guedens:2012sz}. Horizon slices were then assumed to have an entropy density whose form resembles the form of Noether charge conjugate to diffeomorphisms. This was called the Noetheresque entropy. By imposing the Clausius relation on a small patch of the horizon enclosed between two slices sharing a common boundary, it was  shown that
 the field equations for a wide class of higher curvature theories of gravity can be derived if a given consistency condition holds. Unfortunately, this  consistency condition is not satisfied for general theories of gravity containing derivatives of Riemann tensor. Therefore the thermodynamic derivation of field equation is expected to fail in general higher derivative theories of gravity.
Can it be salvaged?

One might wonder why should the entropy density be of the Noetheresque form at all. Could one come up with another definition of entropy of the local causal horizon such that the field equation can be derived from the Clausius relation? Or could one use the ambiguities in the construction of diffeomorphism Noether charge in order to get an entropy that does the job? 
Even in theories without the derivatives of curvature there is a lingering question: how does one define the heat-flux when the matter is non-minimally coupled to the metric? For in that case, there is no canonical splitting of the total Lagrangian between the gravitational part and the matter part. Therefore there is no canonically defined stress tensor that can be used to define the energy flow across the horizon appearing on the right hand side of the Clausius relation.
 
Our goal in this paper is to propose an entropy density that would lead to the derivation of the field equation as an equation of state for any diffeomorphism invariant metric theory of gravity.\footnote{The kind of theories of our interest are those discussed by Iyer and Wald in ref.~\cite{Iyer:1994ys}}. More precisely, we assume that we have a Lagrangian description of a diffeomorphism invariant theory, and we  construct an entropy density associated to slices of the local causal horizon such that imposing the Clausius relation yields the equation of motion of the theory. We will define the heat flux on the right hand side of the Clausius relation by using the stress tensor for a probe field minimally coupled to the metric that we put to zero at the end. This will allow us to work with the total Lagrangian of the theory irrespective of the minimal/non-minimal nature of the matter coupling thus evading  the lingering question mentioned above.

This paper is organized as follows: in sec.~\eqref{sec:LCH} we review the geometry of local causal horizon and the construction of the approximate Killing vector as given in ref.~\cite{Guedens:2012sz}. In sec.~\eqref{sec:EOS} we review the logical steps leading to the derivation of equation of motion as the equation of state via the Clausius relation. In sec.~\eqref{sec:wald_review} we review the Wald-Iyer derivation of the equation of motion for the most general diffeomorphism invariant theory of gravity. In sec.~\eqref{sec:GJS_ent} we review the Noetheresque entropy proposal of ref.~\cite{Guedens:2011dy}.  In sec.~\eqref{sec:DLM_ent} we propose our entropy density and we show that that it leads to the equation of motion, via Clausius relation, for any diffeomorphism invariant metric theory of gravity.
 Some examples are discussed in sec.~\eqref{sec:applications}.  We conclude by presenting the summary and outlook in sec.~\eqref{sec:discussion}.

Our conventions are that of ref.~\cite{Wald:1984rg}. In particular, metric signature is mostly plus and Riemann tensor
 is defined as $2\n_{[a} \n_{b]}\o_c = R_{abc}{}^d \o_d$. After section

\section{Geometry of local causal horizon}
\label{sec:LCH}
There are three essential ingredients involved in the construction of local spacetime thermodynamics. First,  definition of the co-dimension three surface, called the local causal horizon (LCH), which plays the role of the local Rindler horizon.
 Second, specification of a special observer that measures the entropy and the energy flux. Since a general  spacetime has no symmetries there is no Killing vector playing the role of the Rindler observer. Therefore one needs to construct a vector field $\xi$ that is ``approximately'' Killing and plays the role of local observers in whose frame one formulates the local thermodynamics. 
The third and the final ingredient is the specification of the entropy functional associated with the slices of the LCH. In this section we provide a  review of the first two ingredients based on refs.~\cite{Guedens:2011dy,Guedens:2012sz}. The third ingredient, which is also the focus of this paper, will be reviewed in section~\eqref{sec:GJS_ent}.

Let us start with the definition of LCH. Consider a spacelike codimension-two surface $\Sigma_p$ passing through a spacetime point
$p$. This surface has four congruences of null geodesics emanating orthogonally from it: future-pointing and outgoing, future-pointing and ingoing, past-pointing and outgoing, past-pointing and ingoing. The boundary of the past of $\Sigma_p$ has two components generated by the latter two congruences. Pick one of those past boundary components, for concreteness, say, the ingoing one, 
then  our LCH is defined as a small patch of this ingoing past boundary component centered at the point $p$.

\begin{figure}[htb]
\includegraphics[scale=0.6]{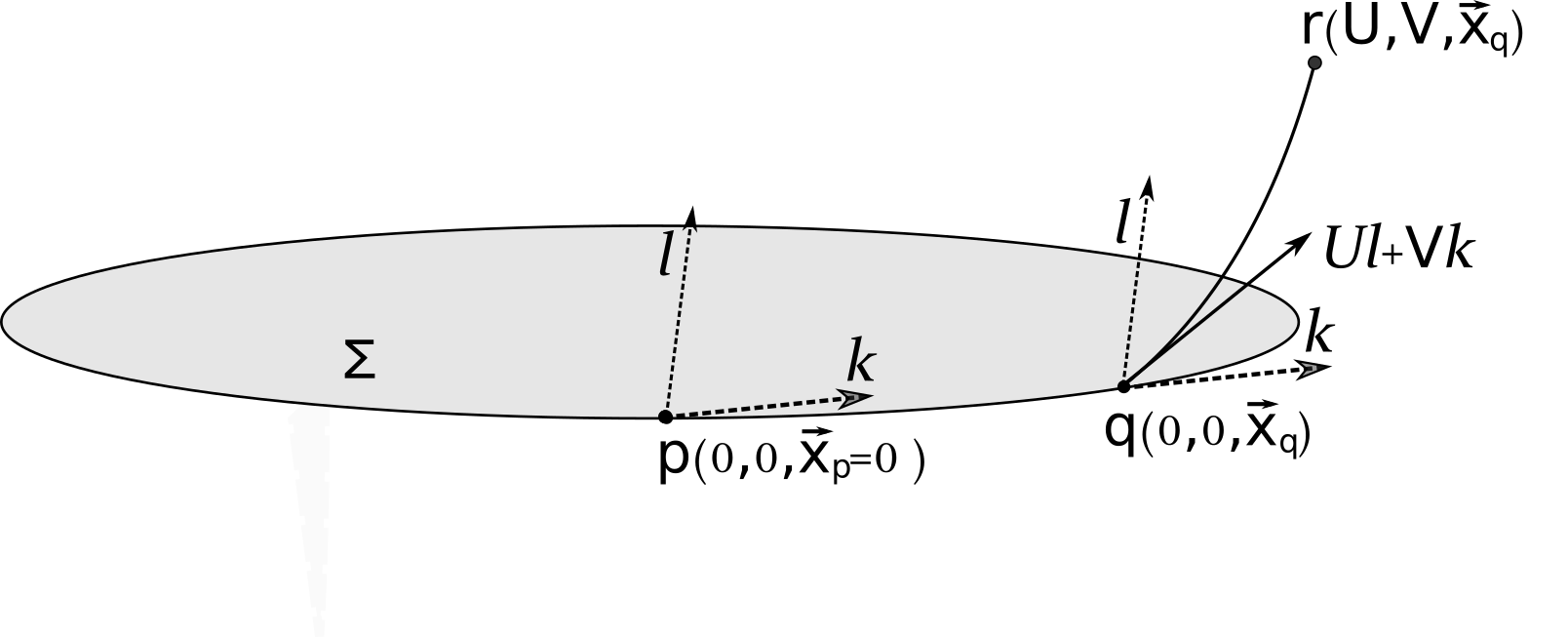}
\caption{ The point $p$ lies on  a $D-2$ dimensional surface  $\Sigma$ which is coordinatized by Riemann Normal Coordinates based upon $p$. Two null vectors $l^a = (\p/\p U)^a$ and $k^a=(\p/\p V)^a$ form the basis of the plane normal to $\Sigma$.  The points off the surface $\Sigma$, say $r$, is coordinatized in terms of the geodesic from $r$  to the surface $\Sigma$ that meets $\Sigma$ orthogonally at the point $q$. If $q$ has coordinates $\{0,0,\vec{x}_q\}$ and the tangent to the geodesic at $q$ is $Vk^a+Ul^a$, then the coordinates of $r$ are $\{U,V,\vec{x}_q\}$. \label{fig:LCHchart}}
\end{figure}

 In order to construct the approximate vector field $\xi$ we choose a coordinate system adapted to LCH 
(see fig.~\ref{fig:LCHchart}). On $\Sigma_p$ we pick the Riemann-normal-coordinates (RNCs) based at $p$. The tangent space orthogonal to the tangent plane of $\Sigma_p$ is spanned by two future-pointing null normals, $\ell^a$ and $k^a$, with the normalization chosen as $\ell^a k_a = -1$. The points off $\Sigma_p$ can then be coordinatized in terms of geodesics orthogonal to $\Sigma_p$ and the points on $\Sigma_p$ where the geodesics emanate from. 
To be more precise, for the point $r$ in the neighborhood of $p$ that lies at the unit affine parameter of the unique geodesic emanating from point $q$ lying  on $\Sigma_p$ is assigned the  coordinates $(U,V,\vec{x}_q)$. Here $\vec{x}_q$ are the RNCs of $q$ and $U,V$ are such that the geodesic running  from $q$ reaching $r$ emanates $q$ in the direction $V k^a + U \ell^a$ orthogonal to $\Sigma_p$. It was shown in ref.~\cite{Guedens:2012sz} that  $k^a$ and $\ell^a$ can be chosen such that the coordinates as defined above are inertial at $p$. The coordinates thus defined are called as  Null Normal Coordinates (NNCs). In these coordinates, LCH is at $U = 0, V \le 0$. 

Next, the approximate Killing vector, called the local Killing vector in ref.~\cite{Guedens:2011dy}, is constructed such that it has a bifurcation point $p_0$ lying on the bifurcation surface $\Sigma_{0}$ to the past of $p$  where it vanishes and where its action is that of a boost in the plane orthogonal to the bifurcation surface. The null generator connecting $p$ to $p_0$ will be called the central generator and will be  denoted by $\G$. In the NNC system $p_0$ has coordinates $ (0,V_0,0)$. 
 It was shown in ref.~\cite{Guedens:2012sz} that an approximate Killing vector $\xi$ with the following properties can be constructed:
\bea
 \xi^\mu \vert_\G &=& (V-V_0) \, \d^\mu_V,  \label{eq:KVprop1} \\  
 \n_{(\mu} \xi_{\nu)} &=& O(x^2),  \label{eq:KVprop2} \\
 \n_\mu \n_\nu \,\xi_\rho |_\G &=&( \left.R_{\rho \nu \mu}{}^{\eta} \, \xi_{\eta})\right\vert_\G.  \label{eq:KVprop3} 
\eea
On the central generator $\G$ local Killing vector is proportional to the generator, $\xi^a = (V-V_0) k^a$. We also have that on $\G$, $\xi^a$ satisfies the geodesic equation with the coefficient of non-affinity given by $\kappa = 1$, i.e., 
\bea
\label{eq:KVprop4}
\left. \xi^b \n_b \xi^a \right\vert_\G = \kappa \xi^a, 
\eea
with $\kappa = 1$. 

Equations~\eqref{eq:KVprop1}~\eqref{eq:KVprop2}~\eqref{eq:KVprop3} were used in the approach of ref.~\cite{Guedens:2011dy} for the derviation of equation of motion as the equation of state by associating a Noetheresque entropy to the slices of LCH. In our approach, we will add an extra term to this entropy and eq.~\eqref{eq:KVprop4} will play a crucial role for the equation of state derivation for any diffeomorphism invariant theory of gravity. 

Before moving on to the third and the final ingredient --the entropy functional-- we pause to review the
 equation of state derivation of the equation of motion. This derivation is at the heart of
local spacetime thermodynamics and would also serve to clarify the role of the aforementioned three ingredients.
 It also gives us an opportunity to introduce a conceptual difference from the previous studies: 
we will view the stress tensor $T_{ab}$ as that of a probe field minimally coupled to
the metric that we will put to zero at the end.
\section{Equation of motion as the equation of state}
\label{sec:EOS}
\begin{figure}
\includegraphics[scale=0.4]{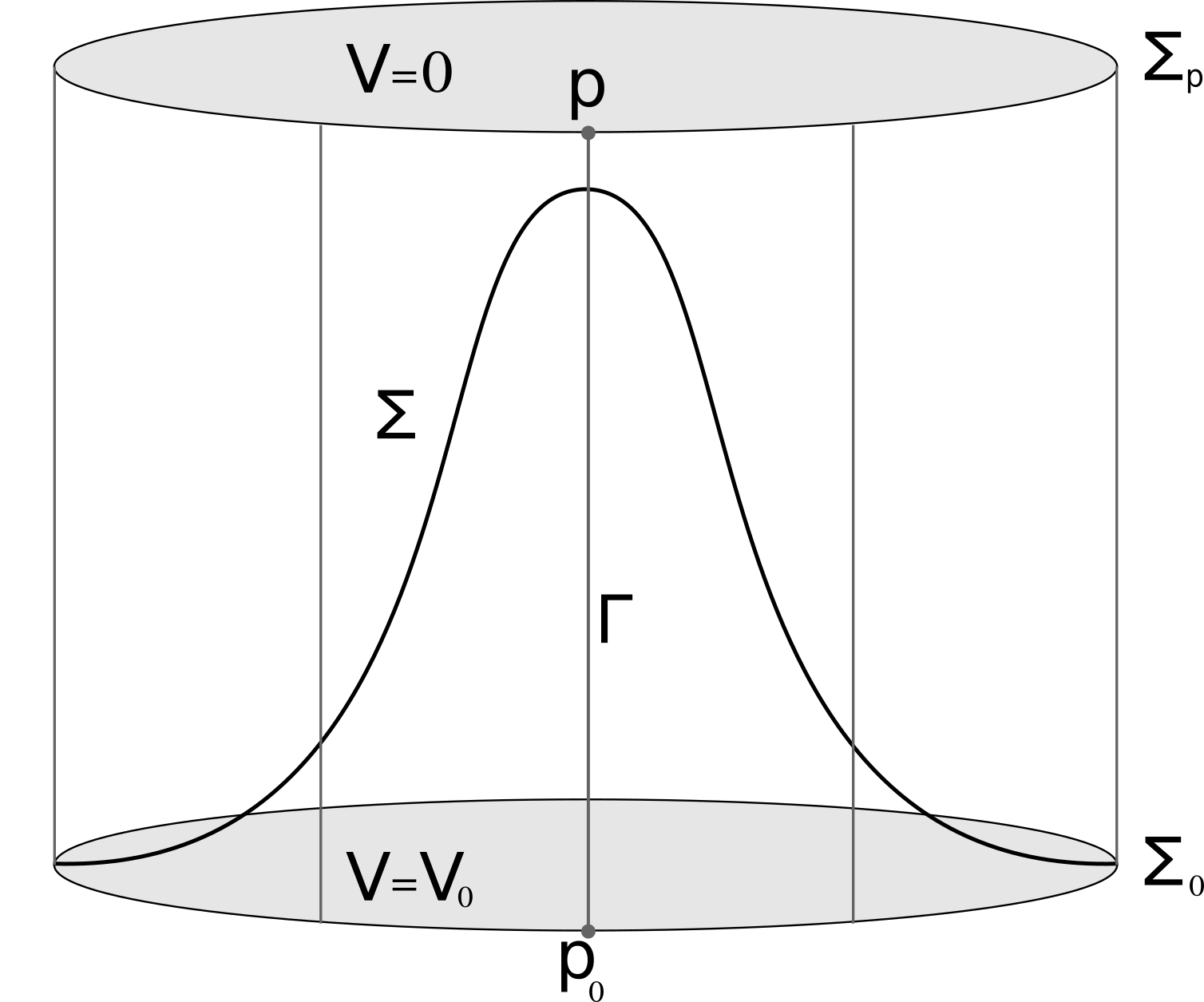}
\caption{The thin narrow patch of LCH surrounding the central generator $\Gamma$ on which the Clausius relation is imposed.\label{fig:LCHpatch}}
\end{figure}

The equation of state derivation of the field equation of a theory of gravity proceeds by imposing the Clausius relation,
\bea
\label{eq:clausius}
\exd S=\frac{\d Q}{T},
\eea
on a thin patch of the LCH, denoted as $\cH$, centered on the central generator $\G$ (see fig.~\ref{fig:LCHpatch}). The left hand side of eq.~\eqref{eq:clausius} is the change in entropy as one evolves the slice of the LCH from $\Sigma_0$ to $\Sigma$ such that they have a common boundary. The right hand side of eq.~\eqref{eq:clausius} contains the temperature, which we choose to have the Unruh value $T=\hbar/2\pi$, and  the heat flux across the patch as measured by the local Killing observer $\xi^a$,
\bea
\label{eq:heatflux}
\d Q = \int_{\cH} (-T_a{}^b \, \xi^a) \,k_b \exd V \,\exd A,
\eea
where $T_{ab}$ is matter energy-momentum tensor,
and the integral is over the thin patch of LCH (see fig.~\ref{fig:LCHpatch}) with the integration measure $ k_a \exd V \exd A$, and $\exd A$ being the volume element on the cut of $\cH$. The integrand of the heat flux is of $O(x)$ since the approximate Killing vector $\xi$ is of $O(x)$. 

In the literature related to the equation of state derivation of the field equation, the stress tensor above is taken to be that of
 the matter fields in the theory. This implicitly assumes that the matter is minimally coupled to the metric, for only then can one separate the total Lagrangian into  a gravitational part and the matter part, and use the latter to define the canonical stress energy tensor. However, in general theories of gravity non-minimal couplings are allowed and there is no natural split between gravity and matter Lagrangian, and thus no natural stress tensor providing the heat flux. We will overcome this problem by deforming the theory
 with a  probe action. We will introduce a probe field  minimally coupled to the metric whose flow drives the evolution of LCH. In the end, we will  put this probe field to zero. Therefore, in our derivation of the equation of state we will take $T_{ab}$ above to be the stress energy tensor of this probe field, $T_{ab} = - 2 \frac{1}{\sqrt{-g}} \frac{\d S_{\rm probe}}{\d g^{ab}}$, where $S_{\rm probe}$ is the action for the probe field minimally coupled to the metric.

To proceed further one needs to specify the change in entropy on the right hand side of eq.~\eqref{eq:clausius}. Intuition from the thermodynamics of black holes suggests that we associate entropy to the slices of LCH.  Following ref.~\cite{Guedens:2011dy}, let
 $s^{ab}$ denote the entropy density (in the dualized form)  associated to an arbitrary slice of LCH.  Total entropy of a slice $\Sigma$ is then given by the integral 
\bea
S=\int_\Sigma s^{ab} n_{ab}\, \exd A,
\eea
where $n^{ab}$ is binormal to the cut $\Sigma$. 
Hence, the change in entropy between two slices $\Sigma$ and $\Sigma_0$ of LCH is given by 
\begin{eqnarray}
\label{eq:ds}
\exd S&=&\int_{\Sigma \cup \Sigma_0}s^{ab} n_{ab} dA
\nonumber \\
&=&-2\int_{\cH} \nabla_b s^{ab}k_a dV dA,
\end{eqnarray}
where the Stokes' theorem was used in the second step. It is at this step that we used that $\Sigma_0$ and $\Sigma$ have the same boundary.
Now imposing the Clausius relation~\eqref{eq:clausius} in the limit $p_0 \rightarrow p$, we get from eqns.~\eqref{eq:heatflux} and ~\eqref{eq:ds}
\begin{eqnarray}
\label{eq:eos}
- (\hbar/\pi)\nabla_b s^{ab}\,  k_a = T^{ab}\xi_b\, k_a + O(x^2). 
\end{eqnarray}
By equating the $O(x)$ terms on both sides of  eq.~\eqref{eq:eos}  at all points $p$ and for all null vectors $k^a$, if we recover  the field equation of the theory of gravity under consideration (after putting $T^{ab} = 0$ because our $T^{ab}$ is that of the probe field) then we deem the program to derive the equation of motion as the equation of state to be successful. Now it is clear that the last ingredient in this program is the specification of the entropy density $s^{ab}$ such that eq.~\eqref{eq:eos} gives the field equation of the theory. 

 Before moving on to the entropy density we should discuss the actual equation of motion for a general diffeomorphism invariant theory of gravity that  we intend to recover from the Clausius relation.  In the rest of this paper we will put $\hbar/2\pi$ to be equal to $1$, i.e., the Unruh temperature is scaled to unity, which is equivalent to choosing a convenient unit for the entropy density
\section{Equation of motion for a general theory of gravity}
\label{sec:wald_review}
In this section we review the  equation of motion of a general diffeomorphism invariant metric theory of gravity following
 ref.~\cite{Iyer:1994ys}. Lagrangian n-form is denoted in bold as $\bL = \e L$. The most general diffeomorphism invariant Lagrangian is of the form 
\begin{widetext}
\bea
\bL = \bL\left[ g_{ab},R_{abcd},\n_{a_1}R_{abcd},..., \n_{(a_1} ... \n_{a_m)} R_{abcd}, \psi, \n_{a_1} \psi, ...,  \n_{(a_1} ... \n_{a_l)} \psi\right], \nn \\
\eea
\end{widetext}
 where $\psi$ denote the matter fields.

The equation of motion for $g_{ab}$ following from the above Lagrangian is given by,
\bea
\label{eq:genEOM}
 A^{ab} + E^{pqra}R_{pqr}{}^b + 2 \n_p \n_q E^{pabq} = 0, \nn \\
\eea
 where $E^{abcd}$ would be the equation of motion for $R_{abcd}$ if we were to treat it as an independent field,
\begin{widetext}
\bea
E^{abcd} = \frac{\p L}{\p R_{abcd}} - \n_{a_1} \frac{\p L}{\p\n_{a_1}R_{abcd} } + ... + 
(-1)^m \n_{(a_1} ... \n_{a_m)} \frac{\p L}{\p \n_{(a_1} ... \n_{a_m)}R_{abcd}},
\eea 
\end{widetext}
and $A^{ab}$ is 
\bea
\label{eq:A-term}
A^{ab} &=& \frac{\p L}{\p g_{ab}} + \frac{1}{2}g^{ab}L + B^{ab}.
\eea
 The origin of the last term $B^{ab}$ is as follows: a typical term in the variation of the Lagrangian due to the derivatives
of Riemann is of the form
\bea
\e \frac{\p L}{\p \n_{(a_1} ... \n_{a_i)} R_{abcd}} \d  \n_{(a_1} ... \n_{a_i)} R_{abcd}, 
\eea
and this can be calculated as 
\bea
\label{eq:typical-term}
 &=& \e \,\, \frac{\p L}{\p \n_{(a_1} ... \n_{a_i)} R_{abcd}}  \n_{a_1} \d \n_{(a_2} ... \n_{a_i)} 
R_{abcd} \nn  \\
&+& \e  \,\, \cdot (\text{terms proportional to} \,\,\n \d g)   \nn \\
&=& \text{exact differential} \nn\\
&+& \text{terms contributing to}\,\, E_{abcd} \nn \\
&+& \e \,\, \cdot (\text{terms proportional to}\,\, \d g),
\eea
where integration by parts was used in both the terms in going from the first equality to the second equality. It is the last term of eq.~\eqref{eq:typical-term}, which is proportional to $\d g_{ab}$, that we denoted as  $B^{ab}$ appearing as the last term in eq.~\eqref{eq:A-term}. We direct the reader to ref.~\cite{Iyer:1994ys} for the details. 
Let us note here that the equation of motion~\eqref{eq:genEOM} can not in general be split in the form  ``$geometry = matter$'' because the total Lagrangian in general does not allow such a split unambiguously. This is the reason why in our equation of state derivation we have to resort to a probe field stress tensor.  

In the next section we will review the Noetheresque entropy proposal of ref.~\cite{Guedens:2011dy} for $s^{ab}$. We will see that while this entropy is  able to derive the equation of motion for theories containing no derivatives of the Riemann tensor, it does not work for theories containing derivatives of Riemann. In sec.~\eqref{sec:DLM_ent} we propose a new entropy density that we will use to derive the equation of motion as the equation of state for any diffeomorphism invariant metric theory of gravity.
\section{Review of the Noetheresque entropy density} 
\label{sec:GJS_ent}
A specific proposal for the entropy density $s^{ab}$ was made in ref.~\cite{Guedens:2011dy} (see also, refs.~\cite{Parikh:2009qs,Padmanabhan:2009ry}). Taking clue from the Noether charge entropy in black hole thermodynamics  ref.~\cite{Guedens:2011dy} proposed 
a Noetheresque form for the entropy density,
\begin{eqnarray}
\label{eq:Ted_ent}
s^{ab}  &=& W^{abc} \xi_c + X^{abcd} \nabla_{[c} \xi_{d]}, 
\end{eqnarray} 
where the tensors $W$ and $X$  are theory dependent quantities and $X$ is antisymmetric in the last two indices. One could also add a term proportional to the symmetric derivative of $\xi$ but it can be shown using the properties~\eqref{eq:KVprop2} and ~\eqref{eq:KVprop3} of the approximate Killing vector that such a term contributes at $O(x^A)$ (where $x^A$ is transverse coordinate in NNC system) to the divergence of entropy density and hence does not contribute to $\d S$ when integrated over small and narrow horizon patches \cite{Guedens:2011dy}.

Calculating the divergence of entropy density~\eqref{eq:Ted_ent} we get,
\bea
\label{eq:divs}
\nabla_b s^{ab}   &=& \left( \nabla_p W^{aps} + X^{apqr} R_{rqp}{}^{s} \right) \xi_s \nn \\
& + & X^{apqr} \left( \n_p \n_q \xi_r  - R_{rqp}{}^s\, \xi_s \right) \nn \\
& + & \left( W^{apq} + \n_r X^{arpq} \right) \n_p \xi_q.
\eea
In this equation the first term is $O(x)$, the second term is $O(x^A)$ due to the Killing identity of eq.~\eqref{eq:KVprop3}, and the third term has an $O(x^2)$ term due to the approximate Killing equation ~\eqref{eq:KVprop2}  and an $O(1)$ term due to the antisymmetric part of the derivative of $\xi$. Since the heat-flux in eq.~\eqref{eq:heatflux} is of $O(x)$ the latter should vanish. Thus we are forced to impose 
\bea
 W^{a[pq]} + \n_r X^{ar[pq]} = 0.
\eea
Since $W$ is antisymmetric in the first two indices, this equation can be solved for $W$ in terms of $X$ \cite{Guedens:2011dy} as
\bea
\label{eq:W}
W^{apq} =  \nabla_r  \left( X^{rapq} + X^{rqpa} + X^{rpqa} \right).
\eea
Putting this back in the eq.~\eqref{eq:divs}, then substituting  $\n_b s^{ab}$ in eq.~\eqref{eq:eos}, and imposing Clausius relation for all $k^a$ we get, 
\bea
\label{eq:eom}
X^{pqr(a}R_{pqr}{}^{b)} - 2 \n_p \n_q X^{p(ab)q} + \Phi g^{ab} =  - \frac{1}{2} T^{ab}, \nn \\
\eea
where $\Phi$ is a scalar that is a function of metric and curvature. Origin of the factor 1/2 on the right hand side is the 
 convention we adopted at the end of sec.~\eqref{sec:EOS} that  $\hbar/2\pi = 1$.
Comparing eq.~\eqref{eq:eom} with the equation of motion for a general diffeomorphism invariant theory eq.~\eqref{eq:genEOM} we see that in general there is no choice of $X$ that would make 
them identical. 

We now recall that in refs.~\cite{Guedens:2011dy,Parikh:2009qs,Padmanabhan:2009ry}  matter was assumed to be minimally coupled, i.e., total Lagrangian $L$ was the the sum of gravitational part and the minimally coupled  matter part $L = L_{(gr)} + L_{(m)}$, and the gravitational part $L_{(gr)}$ was assumed to depend only on the metric and its curvature but not on the derivatives of curvature. Furthermore, the heat flux in the Clausius relation was sourced by the matter stress energy tensor,
\bea
\label{eq:mSTE}
\frac{1}{2} T_{(m)}^{ab} = \frac{\p L_{(m)}}{\p g_{ab}} + \frac12 L_{(m)} g^{ab}.
\eea
Now choosing $-X^{abcd}=  {\p L_{(gr)}}/{\p R_{abcd} }  \equiv P^{abcd}$, and $\Phi = 1/2 L_{(gr)}$ in eq.~\eqref{eq:eom} we get,
\bea
\label{eq:eomXE}
-P^{pqr(a}R_{pqr}{}^{b)} + 2 \n_p \n_q P^{p(ab)q} + \frac{1}{2} L_{(gr)} g^{ab} =  -\frac{1}{2} T_{(m)}^{ab}. \nn \\
\eea  
If the gravity Lagrangian $L_{(gr)}$  does not contain the derivatives  of Riemann then 
there exists an interesting identity,
\bea
\label{eq:Paddy}
\frac{\p L_{(gr)}}{\p g_{ab}} = - 2 P^{pqr(a}R_{pqr}{}^{b)}.
\eea
 This identity, first derived in ref.~\cite{Padmanabhan:2011ex}, is reviewed in app.~\eqref{app:consLLI} where we slightly generalize by considering the gravity Lagrangians containing upto one derivative of curvature.  
Substituting the identity~\eqref{eq:Paddy} in eq.~\eqref{eq:eomXE}, plugging in the expression for $T_{(m)}^{ab}$ from eq.~\eqref{eq:mSTE} and bringing it to the left hand side, we get 
\bea
\label{eq:highercurvEOM}
\frac{\p L}{\p g_{ab}} + P^{pqr(a}R_{pqr}{}^{b)} + 2 \n_p \n_q P^{p(ab)q} + \frac{1}{2} L g^{ab} =  0, \nn \\
\eea
where we have combined the contributions of $L_{(m)}$ and $L_{(gr)}$ into that of the total Lagrangian $L$. Now noticing that  ${\p L_{(gr)}}/{\p R_{abcd} }  \equiv P^{abcd} ={\p L}/{\p R_{abcd} } $ since the matter is minimally coupled, we find that eq.~\eqref{eq:highercurvEOM} is identical  to eq.~\eqref{eq:genEOM} since for higher curvature theories without the derivatives of curvature we have that $B^{ab}=0$,  $A^{ab} = {\p L}/{\p g_{ab}} + {1}/{2} L g_{ab}$ and $E^{abcd} = P^{abcd}$.

Therefore we see that for higher curvature gravity the Noetheresque 
entropy~\eqref{eq:Ted_ent}  of ref.~\cite{Guedens:2011dy} reproduces the equation of motion via the Clausius relation.
  However, for the theories containing derivatives of curvature the equation of motion~\eqref{eq:eom} obtained from the Clausius relation, assuming the Noetheresque entropy as in eq.~\eqref{eq:Ted_ent}, is not the same as the equation of motion of the theory~\eqref{eq:genEOM}. The difference can be traced back as due to the presence of two terms in $A^{ab}$~\eqref{eq:A-term} appearing in the equation of motion: first is ${\p L}/{\p g_{ab}}$, and the second is that arising from the variation $\d  \n\dots\n (\text{Riem})$ of derivative(s) of curvature terms in the Lagrangian that we have collectively denoted as $B^{ab}$. 
 
 In the next section we propose a new definition of entropy that takes care of the uncompensated terms and yields the equation of motion via the Clausius relation. We will view the heat flux  on the right hand side of the Clausius relation as due to the  $T^{ab}$ of a probe field that we will put to zero at the end of the calculation. 
\section{New proposal for the entropy density}
\label{sec:DLM_ent}
In this section we finally present the key finding of this paper. We modify the Noetheresque entropy of eq.~\eqref{eq:Ted_ent} by adding a term quadratic in the approximate Killing vector. Let us introduce a symmetric tensor $M^{ab}$, which will be fixed later depending upon the theory, and consider the following entropy density,
\bea
\label{eq:DLM}
s^{ab}  &=& W^{abc} \xi_c + X^{abcd} \nabla_{[c} \xi_{d]} + 2 M^{c[a}\xi^{b]} \xi_c.
\eea
The $M$ term we have added is of $O(x^2)$ but it contributes at $O(x)$  to the left hand
side in eq.~\eqref{eq:eos}. Let us calculate the divergence of the $M$ term,
\bea
\label{eq:divM}
&\n_b&( 2 M^{c[a}\xi^{b]} \xi_c) \nn \\
 &=& 2(\n_b M^{c[a})\xi^{b]} \xi_c + 2 M^{c[a}(\n_b\xi^{b]}) \xi_c \nn
                               + 2 M^{c[a}\xi^{b]} \n_b\xi_c.
\eea
Here, the first term on the right hand side is of $O(x^2)$. The second term, upon opening the antisymmetrization, has two sub-terms: the first containing $\n_b \xi^b$ is of $O(x^3)$, while the second containing $\n_b \xi^a$ will give zero when contracted with $k_a$. This is so because the approximate Killing vector $\xi$ is proportional to $k$ on the central generator $\Gamma$. The third term in eq.~\eqref{eq:divM} again has two sub-terms: the first one with the free index $a$ on $M$ gives $M^{ca} \xi_c$ after using eq.~\eqref{eq:KVprop4}, while the second with free index $a$ on $\xi$ will give zero after contracting with $k$. Therefore, the only contribution of the $M$ term is to add the tensor $M^{ab}$  to the first line of eq.~\eqref{eq:divs}. The relation between $X$ and $W$ as determined
 in eq.~\eqref{eq:W} remains the same. Thus the equation of motion obtained by imposing Clausius relation with the entropy~\eqref{eq:DLM} is
\bea
\label{eq:DLMeom}
X^{pqr(a}R_{pqr}{}^{b)} - 2 \n_p \n_q X^{p(ab)q} + \Phi g^{ab} + M^{ab} =  -\frac{1}{2} T^{ab}, \nn \\
\eea
where $T^{ab}$ is the stress tensor of the probe field. Now, for a given theory of gravity we can simply choose $M^{ab}$ such that eq.~\eqref{eq:DLMeom} is the equation of motion for
the theory (after putting the probe stress tensor on the right hand side to zero). Comparing with the equation of motion of a general theory of gravity eq.~\eqref{eq:genEOM} we see that we could choose
\bea
 X^{abcd} &=& - E^{abcd}, \label{eq:matching1} \\
  M^{ab} &=& \frac{\p L}{\p g_{ab}} + 2E^{pqr(a}R_{pqr}{}^{b)} + B^{ab}, \label{eq:matching2} \\
\Phi &=& \frac{1}{2} L. \label{eq:matching3}
\eea
Actually, the equation of motion only determines the combination $ \Phi g^{ab} + M^{ab}$. Once we have specified $M^{ab}$ then $\Phi$ can be determined by the Bianchi identity. For the choice of $M^{ab}$ that we have made above, by comparing with the actual equation of motion we already know that $\Phi$ should be equal to $1/2 L$ up to a constant. Since $M^{ab}$ is what appears in the expression of  horizon entropy we see that the  entropy is not unique, for the terms proportional to $g^{ab}$ in $M^{ab}$ could equally well be lumped into $\Phi$.   
\section{Examples}
\label{sec:applications}
Our approach so far has been very general. The use of probe field and the addition of a term quadratic in the local Killing vector to entropy density allowed us to give a thermodynamic derivation of the field equation for  a  general theory of gravity. In this section we illustrate our approach in several  examples.
\subsection{General relativity}
As the simplest illustration of our approach let us consider the Einstein-Hilbert Lagrangian with the matter minimally
coupled to the metric. The total Lagrangian is $L = L_{(EH)} + L_{(m)}$, where $L_{(EH)}=R$ and $L_{(m)}$ is the minimally coupled matter Lagrangian.
The coefficients appearing in the entropy density~\eqref{eq:DLM}, as defined in 
eqns.~(\ref{eq:matching1} \ref{eq:matching2} \ref{eq:matching3}), can be calculated to be,
\bea
X^{abcd}&=&-\frac{1}{2}(g^{ac}g^{bd}-g^{ad}g^{bc}), \nn \\
M^{ab} &=& \frac{\p L}{\p g_{ab}} + 2 R^{ab} =  \frac{\p L_{(m)}}{\p g_{ab}} , \nn \\
\Phi &=& \frac{1}{2} L = \frac12 R + \frac12 L_{(m)}, \nn
\eea
and $W^{abc} = 0$, and where in second equality of the $M$ term we used that $\p R/\p g_{ab} = -2 R^{ab}$. The equation implied by the Clausius relation~\eqref{eq:DLMeom} is then
\bea
-R^{ab} + \frac12 (R + L_{(m)})g^{ab} +  \frac{\p L_{(m)}}{\p g_{ab}} = - \frac12 T^{ab}, 
\eea 
where $T^{ab}$ on the right hand side is the stress tensor of the probe.
For vanishing probe, recognizing that $ \p L_{(m)}/\p g_{ab} + 1/2 L_{(m)}g^{ab} = 1/2\, T_{(m)}^{ab}$ is the matter
stress energy tensor, we get the Einstein field equation (in the units such that $16 \pi G = 1$),
\bea
R^{ab} - \frac12 R g^{ab} = \frac12  T_{(m)}^{ab}. \nn 
\eea
This example illustrates explicitly  that the matter Lagrangian, even if minimally coupled, makes a contribution
to the entropy associated with the slices of LCH because of the $M$ term. Therefore our entropy is different
from that of ref.~\cite{Guedens:2011dy} even for the simplest possible case of general relativity. 
\subsection{Dilaton gravity}
 The second example that we consider is a model in two dimensions: a non-minimally coupled  dilaton $\varphi$ with coupling constant $\lambda$ and a Tachyon $T$, given by the action,
\bea
\label{eq:action-dilaton}
S= \int \exd^2 x  \sqrt{-g}\, e^\varphi (R + (\n \varphi)^2 - (\n T)^2 + \mu^2 T^2 + \lambda ). \nn \\ 
\eea  
The black hole solutions in this model were studied in ref.~\cite{Frolov92} and it was shown that black hole physics in general relativity have counterparts in these two-dimensional models. In particular, black hole entropy of charged black holes in this theory was shown to be proporional to $e^{\varphi_H}$, where $\varphi_H$ is the value of dilaton on the horizon. This result can also be obtained from the Noether charge method (see ref.~\cite{Iyer:1994ys}).
The field equation obtained from the action~\eqref{eq:action-dilaton} is 
\bea
\label{eq:eom-dilaton}
\n^a \n^b \varphi + \n^a T \n^b T + g^{ab}\Big( \Big.&-&\frac12 (\n \varphi)^2-\Box \varphi 
 - \frac12 (\n T)^2 \nn \\ &+& \frac{\mu^2 T^2}{2} + \frac{\lambda}{2}\Big.\Big) = 0.
\eea 
There does not seem to be a natural way to write this equation in terms of separate contributions from geometry
and matter. That is, it is not clear how to decompose
the action~\eqref{eq:action-dilaton} into gravitation and matter piece. Therefore,
we do not know what stress tensor should be used to calculate the heat flux. We could
use the Tachyon stress tensor for this purpose but there does not seem to be a good
justification for doing that.

According to the idea  pursued in this paper, we use the whole
Lagrangian to contribute to the entropy while the heat flux is to be determined
by a probe field that we put to zero at the end. Then the field equation~\eqref{eq:eom-dilaton}
can be obtained by assigning entropy density~\eqref{eq:DLM} to LCHs with the coefficient tensors
given by:
\bea
X^{abcd}&=& -\frac{1}{2} e^\varphi (g^{ac}g^{bd}-g^{ad}g^{bc}), \nn \\
M^{ab} &=& e^\varphi\left( - \n^a \varphi \n^b \varphi + \n^a T \n^b T\right), \nn \\
\Phi &=& \frac12 L, \nn 
\eea
where $L$ is the total Lagrangian for the dilaton theory~\eqref{eq:action-dilaton}. Notice
that $X^{abcd}$ corresponds to the black hole entropy.
In this example we have a non-zero $M^{ab}$ not because of the higher derivative terms (there
are none) but because of the non-minimal coupling of the matter. Even if we were
to define the heat flux not by our probe field but by using the stress tensor of the Tachyon $T$, there
would still be  non-trivial contributions to $M^{ab}$ and therefore this term is needed in the entropy density
to get the field equation from local thermodynamics.  
\subsection{Higher curvature gravity}
Let us now consider higher curvature gravity with minimally coupled matter field for which the Noetheresque entropy
of ref.~\cite{Guedens:2011dy} also gives the field equation via the Clausius relation. For the total Lagrangian
given by
\bea
L = L(g_{ab},R_{abcd},\psi,\n_a \psi),
\eea 
the field equation is
\bea
 \frac{\p L}{\p g_{ab}} + \frac{1}{2}g^{ab}L + P^{pqra}R_{pqr}{}^b + 2 \n_p \n_q P^{pabq} = 0, \nn \\
\eea
where $P^{abcd}={\p L}/{\p R_{abcd} }$. The coefficient tensors in our entropy density~\eqref{eq:DLM} are
given by 
\bea
X^{abcd}&=&-P^{abcd}, \nn \\
 M^{ab} &=& \frac{\p L}{\p g_{ab}} + 2P^{pqr(a}R_{pqr}{}^{b)}, \nn \\
\Phi &=& \frac12 L, \nn
\eea
and $W^{abc}$ is given by eq.~\eqref{eq:W}. This should be contrasted with the entropy
density of  ref.~\cite{Guedens:2011dy} that we reviewed in sec.~\eqref{sec:GJS_ent} where $X$ and $\Phi$ were defined by only
the gravitational part of the Lagrangian and there was no $M$ term. If we allow for the non-minimal coupling in the higher derivative gravity then our approach of using the probe stress tensor to define the heat flux and the new entopy density will continue to yield the field equation via the Clausius relation.   
\subsection{ $S=\int \sqrt{-g}f(\Box R)+S_{matter}$}
As a final example we consider  a higher derivative theory with matter minimally coupled to the metric. The gravitational part of the Lagrangian is a general function of $\Box R$ that we denote by $f(\Box R)$. Some special cases of these theories were studied in ref.~\cite{Hindawi:1995cu} to show their equivalence to general relativity coupled to matter fields with exotic potentials. The equation of motion of this theory is
\bea
&&\nabla^a\nabla^b\square f'-\square f' R^{ab}+\nabla ^a f' \nabla^b R-\frac{1}{2}g^{ab}\nabla_cf'\nabla^cR
\nn\\
&&-g^{ab}\square^2f'+1/2fg^{ab}=\frac{1}{2}T_{(m)}^{ab}, 
\eea
where  $f'={\p f(\square R)}/{\partial \square R}$ and $T_{(m)}^{ab}$ is the canonical stress energy tensor determined by the matter action, $\sqrt{-g} \, T_{(m)}^{ab} = 2 \d S_{matter}/\d g_{ab}$. 
 Since the matter is minimally coupled we could in principle use it to define the heat flux in the Clausius relation. From the point of view of this paper though we will treat the whole action to contribute to the entropy while the heat flux would be given by the stress tensor of the probe field.

For this theory the coefficients appearing in the entropy density~\eqref{eq:DLM}, as defined in 
eqns.~(\ref{eq:matching1} \ref{eq:matching2} \ref{eq:matching3}), can be calculated to be,
\bea
X^{abcd}&=&-\frac{1}{2}(g^{ac}g^{bd}-g^{ad}g^{bc}) \Box f',
\nonumber\\
M^{ab}&=&-\frac{1}{2}T_{(m)}^{ab}+\nabla ^a f' \nabla^b R-g^{ab}\square^2f'-\frac{1}{2}g^{ab}\nabla_cf'\nabla^cR,
\nn\\
\Phi &=& \frac{1}{2}f,   \nn  \label{boxfquantities}
\eea
and $W^{abc}$ is given by eq.~\eqref{eq:W}. Had we considered the heat flux to be sourced by the matter instead of the probe field then $T_{(m)}^{ab}$ would have appeared 
on the right hand side of the Clausius relation and would not have apppeared in $M^{ab}$. As we mentioned at the end of 
sec.~\ref{sec:DLM_ent} the terms proportional to $g^{ab}$ in $M^{ab}$ could be absorbed in $\Phi$. But there would still be left the second term $\n^a f' \n^b R $ in $M^{ab}$. This is precisely the type of term whose origin lies in the derivatives of curvature in the action (as shown in sec.~\ref{sec:wald_review}) and could not be produced by the entropy density of ref.~\cite{Guedens:2011dy}.
Therefore, even in the minimally coupled case and without the use of probe fields, the $M$ term would be needed in the entropy density to yield the correct field equation.
\section{Summary and Outlook}
\label{sec:discussion}
In this paper we have proposed a new  expression for the entropy associated to the slices of local causal horizon, eq.~\eqref{eq:DLM}, such that the Clausius relation imposed on a patch of the horizon implies the field equation of the theory under consideration. The theory in question could be any diffeomorphism invariant metric theory of gravity. In order to achieve this result we introduced two new ingredients: first, the heat flux in the Clausius relation is provided by a minimally coupled probe field that we put to zero in the end, and second, the entropy has a new term quadratic in the approximate Killing vector. Let us discuss these two inputs one by one. 

The reason to introduce the probe matter providing the heat flux is to be able to work with the most general diffeomorphism 
invariant theory. This was done because a general diffeomorphism covariant Lagrangian does not admit a canonical split between a 
gravitational part and a matter part. For example, consider a scalar field $\phi$ in the Lagrangian whose coupling to metric is non-minimal of the form $R^{ab} \n_a \phi \n_b \phi$. If we consider this term as contributing to the matter stress tensor and use it to define the heat flux, then on the left hand side of the equation of motion~\eqref{eq:genEOM} we will not include its contribution to $E^{abcd}$. The resulting entropy density will however not match with the black hole entropy in the theory which is determined
by the $E^{abcd}$ of the total Lagrangian by the Walds' formula.  Alternatively, we could count this term as ``gravitational" and use only the canonical kinetic term of $\phi$ to define the heat flux. This would be a viable option, but it does not appear to be a very natural thing to do. On the other hand, the approach of characterising a system completely by perturbing it with probe fields and observing its response is ubiquitous in physics. 
 In short, the need to work with complete generality, and  the compatibility with black hole thermodynamics led us to define the heat flux in Clausius relation by a probe, minimally coupled matter field that we put to zero in the end. If we were not to use the probe fields to define the heat flux, then we would have to restrict to only the theories with  minimally coupled matter field. But even then the new $M$ term in the entropy density would still be needed to derive the field equation of higher derivative gravity. 

We explained the need for extra term(s) in the entropy explicitly in sec.~\eqref{sec:GJS_ent}. In ref.~\cite{Guedens:2011dy} an integrability condition was derived and it was argued that these extra terms cannot be Noetheresque in a general theory. Does this mean that our proposed entropy is non-Noetheresque? To answer this, we recall that the effect of adding an exact form $\exd\bsmu$ to the Lagrangian $n$-form is to shift the Noether charge from $\bQ$ to $\bQ + \xi \cdot \bsmu$ (see ref.~\cite{Iyer:1994ys}). It is easy to check that if we choose the $(n-1)$-form $\bsmu_{a_1...a_{n-1}} = \boldsymbol{\e}_{a_1...a_{n-1}p} M^{pq} \xi_q$  then the Noether procedure will reproduce our proposed additional term in the entropy~\footnote{We thank Ted Jacobson for making this suggestion and the related discussion.}. Therefore, our entropy is also Noetheresque. This is consistent with the result in ref.~\cite{Guedens:2011dy} because the surface term we added is not constructed from just the dynamical fields in the theory but it also contains $\xi$. If one insists on adding only covariant boundary terms then indeed one cannot obtain a term quadratic in $\xi$ in the Noether charge. 
However, as in ref.~\cite{Guedens:2011dy}, since our entropy depends on the arbitrary choice of the bifurcation point its physical significance remains obscure. This dependence on the approximate Killing vector is also the reason why we have evaded the physical arguments of ref.~\cite{Jacobson:2012yt} that concluded that the corrections to Einstein gravity can not be obtained by a thermodynamic reasoning. 

The new term in the entropy that we have proposed does not alter the black hole entropy because the Killing vector vanishes on the bifurcation surface. Compatibility with black hole thermodynamics is a stringent requirement. Without it, we could have simply taken the whole entropy as given by the quadratic term and chosen $M^{ab}$ to be the equation of motion. But then the black hole entropy in the theory would be zero. The $X$ term in eq.~\eqref{eq:DLM} is thus dictated by the black hole entropy. The $W$ term is necessary for the equation of state argument to go through for the higher curvature theories. For higher derivative theories the  $M$ term in eq.~\eqref{eq:DLM} is needed to get the equation of motion via the Clausius relation.

The generality of our approach seems to suggest that there is no obstacle for the equation of state derivation  for any diffeo-invariant metric theory of gravity, irrespective of whether it is Lorentz invariant or not. In particular, one could then derive local thermodynamics in Lorentz violating theories, e.g., the Einstein-{\AE}ther theory. However, this expectation faces two challenges. First of all, local Lorentz invariance is crucial to associate the Unruh temperature with the local causal horizon.
Second,  the existence of black hole  thermodynamics in such theories is not well-settled yet, and  is  under active investigation \cite{Dubovsky:2006vk,Eling:2007qd,Jacobson:2008yc,Foster:2005fr,Berglund:2012bu,Berglund:2012fk,Cropp:2013sea,Michel:2015rsa,Mohd:2013zca}. Finding a  thermodynamic route to the equation of motion in such theories thus appears to be a premature enterprise at the moment. 

We would like to mention in passing that some authors \cite{Mohd:2013jva,Padmanabhan:2009ry} have taken the converse route to the one taken in this paper. That is, their goal is to understand  if the field equation implies the Clausius relation for an appropriately defined entropy density. It is not too difficult to show that given our entropy one  can follow this program of running the argument backwards to its completion for any diffeomorphism invariant theory of gravity.

Finally, we should point out that the entropy density is highly non-unique. This non-uniqueness is beyond the non-uniqueness pointed out at the end of sec.~\eqref{sec:DLM_ent}. It is easy to write down higher order terms in $\xi$ and its derivatives such that their contribution to the change in entropy of the patch of LCH is just $M^{ab} \xi_b$ for some effective $M^{ab}$. We think that the underlying problem is our completely classical treatment of the fields. We believe  that the correct notion of entropy to be used in any thermodynamic derivation of field equation has to be  quantum mechanical one. This is exemplified by a recent derivation
of the semiclassical Einstein equation by Jacobson that involves an ansatz on the nature of  entanglement entropy of the vacuum ~\cite{Jacobson:2015hqa}. It has recently been pointed out in  ref.~\cite{Carroll:2016lku} that relative entropy is not the right quantity to use on the left hand side of the Clausius relation in the geometric framework used here. It remains to be seen what quantum mechanical measure of entropy is rich enough to encode the dynamics of gravity. Until that is found, our entropy expression \eqref{eq:DLM} seems to serve as a plausible place-holder.
\section{Acknowledgments}
 We thank Alessio Belenchia and Ted Jacobson for illuminating discussions. We also acknowledge the John Templeton Foundation for the supporting grant \#51876.  The research of AM was supported by the National Science Foundation under grant number  PHY-1407744.

\appendix
\section{Derivation of the identity in eq.~\eqref{eq:Paddy} }
\label{app:consLLI}
In this appendix we derive the identity in eq.~\eqref{eq:Paddy}. This identity was first derived in ref.~\cite{Padmanabhan:2011ex} whose treatment we follow here. A slight generalization here is that we consider the Lagrangians containing upto one derivative of curvature, $L = L(g _{ab},R_{abcd},\n_{a_1}R_{abcd})$.  In this section $L$ will stand for pure gravitational Lagrangian that we denoted as $L_{(gr)}$ in the main text. 

Let us consider an infinitesimal diffeomorphism $x^a \rightarrow x^a+\xi^a$ 
generated by a vector field $\xi$. The infinitesimal change in $L$ is given by the Lie derivative of $L$ that can be calculated in two different ways.
 In the  first way,  by  considering the dependence of $L$ on $x^a$ through $g_{ab}$, $R_{abcd}$ and  $ \n_{a_1}R_{abcd}$, we can write
\bea
\mathsterling_{\xi} L&&=\xi^m\nabla_m L=P^{abcd} \xi^m\nabla_m R_{abcd}
 \nn\\
 &&+Z^{a_1abcd}\xi^{m}\n_m\n_{a_1}R_{abcd}+A^{ab}\xi^m \n_m g_{ab},   \nn \\ \label{liederivative1}
 \eea       
 where $A^{ab}=\frac{\p L}{\p g_{ab}}$, $P^{abcd}=\frac{\p L}{\p R_{abcd}}$, and $ Z^{a_1abcd}=\frac{\p L}{\p \n_{a_1}R_{abcd}}$. The ${abcd}$ indices of $P$ and $Z$ are taken to have the symmetries of Riemann.
 
The second way is to consider the infinitesimal variation $\d L$ in $L$ as due to  the variation in $g_{ab}$, $R_{abcd}$ and $\n_{a_1}R_{abcd}$ due to the diffeomorphism. The latter are  given by the corresponding Lie derivatives.
 Thus we have,
\begin{eqnarray}
 &&\mathsterling_{\xi} L=A^{ab} \mathsterling_{\xi}g_{ab}
 +P^{abcd}\mathsterling_{\xi}R_{abcd}  
 +Z^{a_1abcd}\mathsterling_{\xi}\n_{a_1}R_{abcd}.     \nn\\             \label{2ndlie}
 \end{eqnarray}
 Taking into consideration the symmetries of $R_{abcd}$ and $P^{abcd}$, the second term can be calculated as
\bea 
 P^{abcd}\mathsterling_{\xi}R_{abcd}=P^{abcd}\big[\xi^m\nabla_mR_{abcd} +4\nabla_{a}\xi^m R_{mbcd}\big].    \nn\\
 \eea
Similarly, the third term can be calculated as
\bea
 &&Z^{a_1abcd}\mathsterling_{\xi}\n_{a_1}R_{abcd} =Z^{a_1abcd}\big[\xi^m\n_m \n_{a_1}R_{abcd}
 \nn\\
 &&+4\n_{a_1}R_{mbcd}\n_{a}\xi^m +\n_m R_{abcd}\n_{a_1}\xi^m \big].
 \eea
 Plugging these expressions in eq.~\eqref{2ndlie}, 
and  using $\mathsterling_{\xi}g_{ab}=\nabla_a\xi_b+\nabla_b\xi_a$,  we get
 \begin{eqnarray}
 \mathsterling_{\xi} L&=&P^{abcd}\xi^m\nabla_m R_{abcd}
 +Z^{a_1abcd}\xi^{m}\n_m\n_{a_1}R_{abcd} \nn \\
&+& 2\nabla_p\xi_q\Big[ \frac{\partial L}{\partial g_{pq}}+2P^{pabc}R^q{}_{abc} \nn \\
 &+& 2Z^{a_1pabc}\n_{a_1}R^q{}_{abc}+\frac{1}{2}Z^{pabcd}\n^{q}R_{abcd} \Big].
 \end{eqnarray}
Now, we see from eq.~\eqref{liederivative1} that the first two terms on the right hand side are already equal to $\mathsterling_{\xi} L$. This implies, since $\xi$ is arbitrary, that the expression within the brackets must vanish. We thus get the identity 
 \begin{widetext}
 \bea
 \frac{\partial L}{\partial g_{pq}}=-2P^{pabc}R^q{}_{abc}
-2Z^{a_1pabc}\n_{a_1}R^q{}_{abc}-\frac{1}{2}Z^{pabcd}\n^{q}R_{abcd}.  \nn\\   \label{eq:diffeoinvconseq}
 \eea
 \end{widetext}
 For  theories containing no derivatives of curvature, we have that $Z^{pabcd}=0$ and eq.~\eqref{eq:diffeoinvconseq} reduces to the identity~\eqref{eq:Paddy}. 
\bibliography{LCH-arXiv}

\end{document}